\documentclass[12pt,NJP]{iopart}

\usepackage{amssymb}
\usepackage{amsfonts}
\usepackage{amsmath}
\usepackage{graphics}
\usepackage{color}
\usepackage{sidecap}
\usepackage{epsfig}
\usepackage{times}
\usepackage{cite}
\usepackage{etoolbox} 
\usepackage{mathrsfs} 
\usepackage{rotating}
\usepackage{multirow}
\usepackage{bbold}
\usepackage{chngpage} 

\newcommand{\avg}[1]{\ensuremath{\overline{ #1  }  }} 
\newcommand{\ensavg}[1]{\ensuremath{\langle #1  \rangle }} 

\newcommand{\supranode}[1][]{ %
\ifstrequal{#1}{}{\ensuremath{u} }{}
\ifstrequal{#1}{1}{\ensuremath{u} }{}
\ifstrequal{#1}{2}{\ensuremath{v} }{}
\ifstrequal{#1}{3}{\ensuremath{w} }{}
}

\newcommand{\micronode}[1][]{ %
\ifstrequal{#1}{}{\ensuremath{i} }{}
\ifstrequal{#1}{1}{\ensuremath{i} }{}
\ifstrequal{#1}{2}{\ensuremath{j} }{}
\ifstrequal{#1}{3}{\ensuremath{h} }{}
}

\begin{document}


\title{Structure of triadic relations in multiplex networks}

\author{Emanuele Cozzo$^1$, Mikko Kivel\"a$^2$, Manlio De Domenico$^3$, Albert Sol\'e-Ribalta$^3$, Alex Arenas$^3$, Sergio G\'omez$^3$, Mason A. Porter$^{2,4}$, Yamir Moreno$^{1,5,6}$}
\address{$^1$ Institute for Biocomputation and Physics of Complex Systems (BIFI), University of Zaragoza, Zaragoza 50018, Spain \\ $^2$ Oxford Centre for Industrial and Applied Mathematics, Mathematical Institute, University of Oxford, Oxford, OX2 6GG, UK \\ $^3$ Departament d'Enginyeria Inform\`atica i Matem\`atiques, Universitat Rovira i Virgili, 43007 Tarragona, Spain \\ $^4$ CABDyN Complexity Centre, University of Oxford, Oxford, OX1 1HP, UK \\ $^5$ Department of Theoretical Physics, University of Zaragoza, Zaragoza 50009, Spain \\ $^6$ Complex Networks and Systems Lagrange Lab, Institute for Scientific Interchange, Turin, Italy}





\begin{abstract}

Recent advances in the study of networked systems have highlighted that our interconnected world is composed of networks that are coupled to each other through different ``layers'' that each represent one of many possible subsystems or types of interactions. Nevertheless, it is traditional to aggregate multilayer networks into a single weighted network in order to take advantage of existing tools. This is admittedly convenient, but it is also extremely problematic, as important information can be lost as a result. It is therefore important to develop multilayer generalizations of network concepts. In this paper, we analyze triadic relations and generalize the idea of transitivity to multiplex networks. By focusing on triadic relations, which yield the simplest type of transitivity, we generalize the concept and computation of clustering coefficients to multiplex networks. We show how the layered structure of such networks introduces a new degree of freedom that has a fundamental effect on transitivity.  We compute multiplex clustering coefficients for several real multiplex networks and illustrate why one must take great care when generalizing standard network concepts to multiplex networks. We also derive analytical expressions for our clustering coefficients for ensemble averages of networks in a family of random multiplex networks. Our analysis illustrates that social networks have a strong tendency to promote redundancy by closing triads at every layer and that they thereby have a different type of multiplex transitivity from transportation networks, which do not exhibit such a tendency.  These insights are invisible if one only studies aggregated networks.

\end{abstract}



\maketitle

\section{Introduction}

The quantitative study of networks is fundamental for investigations of complex systems throughout the biological, social, information, engineering, and physical sciences \cite{faust,bocca2006,newman2010}.  The broad applicability of networks, and their success in providing insights into the structure and function of both natural and designed systems, has generated considerable excitement across myriad scientific disciplines.  Numerous tools have been developed to study networks, and the realization that several common features arise in a diverse variety of networks has facilitated the development of theoretical tools to study them. For example, many networks constructed from empirical data have heavy-tailed degree distributions, satisfy the small-world property, and/or possess modular structures. Such structural features can have important implications for information dissemination, robustness against component failure, and more.

Traditional studies of networks generally assume that nodes are adjacent to each other by a single type of static edge that encapsulates all connections between them.  This assumption is almost always a gross oversimplification, and it can lead to misleading results and even the fundamental inability to address certain problems.  Most real systems have multilayer structures \cite{Kivela2014Multilayer,Boccaletti2014Structure}, as there are almost always multiple types of ties or interactions that can occur between nodes, and it is crucial to take them into account. For example, transportation systems include multiple modes of travel, biological systems include multiple signaling channels that operate in parallel, and social networks include multiple types of relationships and multiple modes of communication. We will represent such systems using the formalism of \emph{multiplex networks}, which allow one to incorporate multiple types of edges between nodes.

The notion of multiplexity was introduced years ago in fields such as engineering \cite{chang1996,little2002} and sociology \cite{faust,clyde1969,verbrugge1979,coleman1988}, but the discussions included few analytical tools to accompany them. This situation arose for a simple reason: although many aspects of single-layer networks are well understood, it is challenging to properly generalize even the simplest concepts to multiplex networks.  Theoretical developments on multilayer networks (including both multiplex networks and interconnected networks) have gained steam only in the last few years \cite{raissa2009,Mucha2010community,criado2010,buldyrev2010,gao2011,raissa2012,yagan2012,gao2012,lee2012,brummitt2012}, and even basic notions like centrality and diffusion have barely been studied in multiplex settings \cite{gomez2013diffusion,dedomenico2013random,Bianconi2013Statistical,Sola2013Centrality,pagerank2013,cozzo2012,cozzo2013,granell13,sole13,versatility15}.  New degrees of freedom arise from the multilayer structure of multiplex networks, and this brings both new challenges \cite{tensorial13,Kivela2014Multilayer} and new phenomena. The new phenomena include multiplexity-induced correlations \cite{lee2012}, new types of dynamical feedbacks \cite{cozzo2012}, and ``costs" of inter-layer connections \cite{mingoh2013}. For reviews about networks with multiple layers, see Refs.~\cite{Kivela2014Multilayer,Boccaletti2014Structure}.

In the present article, we focus on one of the most important structural properties of networks: triadic relations, which are used to describe the simplest and most fundamental type of transitivity in networks \cite{luce1949,Watts1998Collective,faust,newman2010,karlberg1997}. We develop multiplex generalizations of clustering coefficients, which can be done in myriad ways, and (as we will illustrate) the most appropriate generalization depends on the application under study. Such considerations are crucial when developing multiplex generalizations of any single-layer (i.e., ``monoplex'') network diagnostic.
There have been several attempts to define multiplex clustering coefficients~\cite{Barrett2012Taking,Brodka2010Method,Brodka2012Analysis,Criado2011Mathematical,Battiston2014Metrics}, but there are significant shortcomings in these definitions. For example, some of them do not reduce to the standard single-layer clustering coefficient or are not properly normalized (see \ref{sec:mplexcc}).

The fact that existing definitions of multiplex clustering coefficients are mostly \emph{ad hoc} makes them difficult to interpret. In our definitions, we start from the basic concepts of walks and cycles to obtain a transparent and general definition of transitivity. This approach also guarantees that our clustering coefficients are always properly normalized. It reduces to a weighted clustering coefficient~\cite{Zhang2005General} of an aggregated network for particular values of the parameters; this allows comparison with existing single-layer diagnostics. We also address two additional, very important issues: (1) Multiplex networks have many types of connections, and our multiplex clustering coefficients are (by construction) decomposable, so that the contribution of each type of connection is explicit; (2) because our notion of multiplex clustering coefficients builds on walks and cycles, we do not require every node to be present in all layers, which removes a major (and very unrealistic) simplification that is used in existing definitions.

Using the example of clustering coefficients, we illustrate how the new degrees of freedom that result from the existence of multiple layers in a multiplex network yield rich new phenomena and subtle differences in how one should define key network diagnostics.  As an illustration of such phenomena, we derive analytical expressions for the expected values of clustering coefficients on multiplex networks in which each layer is an independent Erd\H{o}s-R\'{e}nyi (ER) graph. We find that the clustering coefficients depend on the intra-layer densities in a nontrivial way if the probabilities for an edge to exist are heterogeneous across the layers. We thereby demonstrate for multiplex networks that it is insufficient to generalize existing diagnostics in a na\"{\i}ve manner and that one must instead construct their generalizations from first principles (e.g., as walks and cycles in this case).

\section{Methods}
\subsection{Mathematical Representation}
\label{sec:mathematical}

We represent a multiplex network using a finite sequence of graphs $\{G^\alpha \}$, with $G^\alpha=(V^\alpha,E^\alpha)$, where $\alpha \in L$ is the set of layers. Without loss of generality, we let $L=\{1,\ldots,b\}$ and $V^{\alpha} \subseteq \{1,\ldots,n\}$. For simplicity, we examine unweighted and undirected multiplex networks. We define the intra-layer supra-graph as $G_A=(V,E_A)$, where the set of nodes is $V=\bigcup_\alpha \{(\supranode[1],\alpha): \supranode[1] \in V^\alpha \}$ and the set of edges is $E_A=\bigcup_\alpha \{ ((\supranode[1],\alpha),(\supranode[2],\alpha)) : (\supranode[1],\supranode[2]) \in E^\alpha\}$.  We also define the coupling supra-graph $G_C=(V,E_C)$ using the same sets of nodes and the edge set $E_C=\bigcup_{\alpha,\kappa}\{ ((\supranode,\alpha),(\supranode,\kappa)) : \supranode \in V^\alpha , \supranode \in V^\kappa, \alpha\neq\kappa \}$ and its associated adjacency matrix $\mathcal{C}$. If $((\supranode,\alpha),(\supranode,\kappa)) \in E_C$, we say that $(\supranode,\alpha)$ and $(\supranode,\kappa)$ are ``interconnected.'' 
The nonzero entries of the matrix $\mathcal{C}=\mathcal{C}^T$ indicate the connections between corresponding nodes (i.e., between the same entity) on different layers. We say that a multiplex network is ``node-aligned'' \cite{Kivela2014Multilayer} if all layers share the same set of nodes (i.e., if $V^\alpha = V^\kappa$ for all $\alpha$ and $\kappa$). The supra-graph is $\bar{G}=(V,\bar{E})$, where $\bar{E}=E_A \cup E_C$. The corresponding adjacency matrix is the supra-adjacency matrix $\mathcal{\bar{A}}$.

Supra-adjacency matrices satisfy $\mathcal{\bar{A}}=\mathcal{A}+\mathcal{C}$ and $\mathcal{A}=\bigoplus_\alpha \mathbf{A}^{(\alpha)}$, where $\mathbf{A}^{(\alpha)}$ is the adjacency matrix of layer $\alpha$ (i.e., the adjacency matrix associated to $G^\alpha$) and $\bigoplus$ denotes the direct sum of the matrices. We consider undirected networks, so $\mathcal{A}=\mathcal{A}^T$. For clarity, we denote nodes in a given layer
and in monoplex networks using the symbols $\supranode[1], \supranode[2], \supranode[3]$; and we denote indices
in a supra-adjacency matrix using the symbols $\micronode[1], \micronode[2], \micronode[3]$.
We also define $l(\supranode[1])=\{(\supranode[1],\alpha) \in V | \alpha \in L \}$ to be the set of supra-adjacency matrix indices that correspond to node $\supranode[1]$, and we refer to the nodes $(u,\alpha)$ of a supra-graph as a node-layer pair.  An entity $u$ corresponds to a ``physical node.''

The local clustering coefficient $C_{\supranode[1]}$ of node $\supranode$ in an unweighted monoplex network is the number of triangles (i.e., triads) that include node $\supranode$ divided by the number of connected triples (i.e., either 2-stars or triangles) with node $\supranode$ in the center~\cite{Watts1998Collective,newman2010}. The local clustering coefficient is a measure of transitivity \cite{luce1949}, and it can be interpreted as the density of a focal node's neighborhood.  For our purposes, it is convenient to define the local clustering coefficient $C_{\supranode}$ as the number of 3-cycles $t_{\supranode}$ that start and end at the focal node $\supranode$ divided by the number of 3-cycles $d_{\supranode}$ such that the second step of the cycle occurs in a complete graph (i.e., assuming that the neighborhood of the focal node is as dense as possible).\footnote{Note that we use the term ``cycle'' to refer to a walk that starts and ends at the same physical node $\supranode$.  As we will discuss later, in a multiplex network, it is permissible (and relevant) to return to the same node via a different layer from the one that was used originally to leave the node.}
In mathematical terms, $t_{\supranode}=(\mathbf{A}^3)_{\supranode\supranode}$ and $d_{\supranode}=(\mathbf{AFA})_{\supranode\supranode}$, where $\mathbf{A}$ is the adjacency matrix of the graph and $\mathbf{F}$ is the adjacency matrix of a complete graph with no self-edges. (In other words, $\mathbf{F}=\mathbf{J}-\mathbf{I}$, where $\mathbf{J}$ is a complete square matrix of $1$s and $\mathbf{I}$ is the identity matrix.)

The \emph{local clustering coefficient} for node $\supranode$ is thus given by the formula $C_{\supranode}=t_{\supranode}/d_{\supranode}$. This is equivalent to the usual definition of the local clustering coefficient: $C_{\supranode}=t_{\supranode}/(k_{\supranode}(k_{\supranode}-1))$, where $k_{\supranode} \geq 2$ is the degree of node $\supranode$. (The local clustering coefficient is often set to $0$ for nodes of degree $0$ and $1$, and another option is to state that it is not defined in such cases.) One can calculate a single global clustering coefficient for a monoplex network either by averaging $C_{\supranode}$ over all nodes or by computing $C=\frac{\sum_{\supranode} t_{\supranode}}{\sum_{\supranode} d_{\supranode}}$. Henceforth, we will use the term \emph{global clustering coefficient} for the latter quantity.


\subsection{Triads on Multiplex Networks}

In addition to 3-cycles (i.e., triads) that occur within a single layer, multiplex networks also contain cycles that incorporate more than one layer but still have 3 intra-layer steps. Such cycles are important for the analysis of transitivity in multiplex networks. In social networks, for example, transitivity involves social ties across multiple social environments \cite{faust,szell2010}. In transportation networks, there typically exist several means of transport to return to one's starting location, and different combinations of transportation modes are important in different cities \cite{gallotti2014}. For dynamical processes on multiplex networks, it is important to consider 3-cycles that traverse different numbers of layers, so one needs to take them into account when defining a multiplex clustering coefficient. We define a \emph{supra-walk} as a walk on a multiplex network in which, either before or after each intra-layer step, a walk can either continue on the same layer or change to an adjacent layer. We represent this choice using the following matrix:
\begin{equation}
	\mathcal{\widehat{C}} = \beta\mathcal{I} + \gamma\mathcal{C}\,,
\label{Cwithhat}
\end{equation}
where $\mathcal{I}$ is the $|V| \times |V|$ identity matrix, $|V|$ is the number of node-layer pairs, the parameter $\beta$ is a weight that accounts for the walk staying in the current layer, and $\gamma$ is a weight that accounts for the walk stepping to another layer. In a supra-walk, a \emph{supra-step} consists either of only a single intra-layer step or of a step that includes both an intra-layer step and an inter-layer step, in which one changes from one layer to another (either before or after the intra-layer step). In the latter type of supra-step, note that we are disallowing two consecutive inter-layer steps. The number of 3-cycles for node $\micronode$ is then
\begin{align}
	t_{M,\micronode}=[(\mathcal{A\widehat{C}})^3 + (\mathcal{\widehat{C}A})^3]_{\micronode\micronode} \label{taw}\,,
\end{align}
where the first term corresponds to cycles in which the inter-layer step is taken after an intra-layer one and the second term corresponds to cycles in which the inter-layer step is taken before an intra-layer one. The subscript $M$ refers to the particular way that we define a supra-walk in a multiplex network through the supra-matrices $\mathcal{A\widehat{C}}$ and $\mathcal{\widehat{C}A}$. However, one can also define other types of supra-walks (see \ref{sec:othercycles} and \ref{sec:auxiliary}), and we will use different subscripts when we refer to them. We can simplify Eq.~{\bf\ref{taw}} by exploiting the fact that both $\mathcal{A}$ and $\mathcal{\widehat{C}}$ are symmetric.  This yields
\begin{align}
	t_{M,\micronode}=2[(\mathcal{A\widehat{C}})^3]_{\micronode\micronode}\,. \label{tawsimple}
\end{align}

\begin{figure*}[t]
\includegraphics[width=0.99\textwidth]{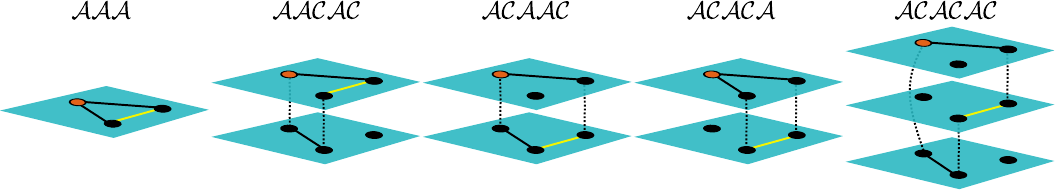}
\caption{Sketch of the elementary cycles $\mathcal{AAA}$, $\mathcal{AACAC}$, $\mathcal{ACAAC}$, $\mathcal{ACACA}$, and $\mathcal{ACACAC}$. The orange node is the starting point of the cycle. The intra-layer edges are the solid lines, and the intra-layer edges are the dotted curves. In each case, the yellow line represents the second intra-layer step.}
\label{fig:elementary_cycles}
\end{figure*}

It is useful to decompose multiplex clustering coefficients that are defined in terms of multilayer cycles into so-called \emph{elementary cycles} by expanding Eq.~{\bf\ref{tawsimple}} and writing it in terms of the matrices $\mathcal{A}$ and $\mathcal{C}$. That is, we write $t_{M,\micronode}=\sum_{\mathcal{E} \in \mathscr{E}} w_{\mathcal{E}}(\mathcal{E})_{\micronode\micronode}$, where $\mathscr{E}$ denotes the set of elementary cycles and $w_{\mathcal{E}}$ are weights of different elementary cycles. We can use symmetries in our definition of cycles and thereby express all of the elementary cycles in a standard form with terms from the set $\mathscr{E} = \{\mathcal{AAA}, \mathcal{AACAC},\mathcal{ACAAC}, \mathcal{ACACA}, \mathcal{ACACAC}\}$. See Fig.~\ref{fig:elementary_cycles} for an illustration of elementary cycles and \ref{sec:elemcycles} for details on deriving the elementary cycles. Note that some of the alternative definitions of a 3-cycle---which we discuss in \ref{sec:othercycles}---lead to more elementary cycles than the ones that we just enumerated.

To define multiplex clustering coefficients, we need both the number $t_{*,\micronode}$ of cycles and a normalization $d_{*,\micronode}$. The symbol $*$ stands for any type of cycle: the 3-cycle that we define in the main text, an elementary cycle, or the alternatives definition of 3-cycles that we give in \ref{sec:othercycles}. Choosing a particular definition coincides to a given way to calculate the associated expression for $t_{*,\micronode}$. To determine the normalization, it is natural to follow the same procedure as with monoplex clustering coefficients and use a complete multiplex network $\mathcal{F}=\bigoplus_\alpha \mathbf{F}^{(\alpha)}$, where $\mathbf{F}^{(\alpha)}=\mathbf{J}^{(\alpha)}-\mathbf{I}^{(\alpha)}$ is the adjacency matrix for a complete graph on layer $\alpha$. We can then proceed from any definition of $t_{*,i}$ to $d_{*,i}$ by replacing the second intra-layer step with a step in the complete multiplex network. For example, we obtain $d_{M,\micronode}=2(\mathcal{A\widehat{C}F\widehat{C}A\widehat{C}})_{\micronode\micronode}$ for $t_{M,\micronode}=2[(\mathcal{A\widehat{C}})^3]_{\micronode\micronode}$. Similarly, one can use any other definition of a cycle (e.g., any of the elementary cycles or the cycles that we discuss in \ref{sec:othercycles}) as a starting point for defining a multiplex clustering coefficient.

The above formulation allows us to define local and global clustering coefficients for multiplex networks analogously to their definition in monoplex networks. We can calculate a natural multiplex analog to the usual monoplex local clustering coefficient for any node $\micronode$ of the supra-graph. Additionally, in a multiplex network, a node $\supranode$ of an intra-layer network allows an intermediate description for clustering that lies between local and the global clustering coefficients. We define
\begin{align}
	c_{*,i} &= \frac{t_{*,i}}{d_{*,i}}\,, \label{clocal} \\
	C_{*,\supranode} &= \frac{\sum_{i \in l(\supranode) } t_{*,i}}{\sum_{i \in l(\supranode)} d_{*,i}}\,, \label{suprac} \\
	C_{*} &= \frac{\sum_i t_{*,i}}{\sum_i d_{*,i}}\,, \label{cglobal}
\end{align}
where $l(\supranode)$ is as defined before. Note that we refer to clustering coefficients defined either by Eq.~{\bf\ref{clocal}} or by Eq.~{\bf\ref{suprac}} as local clustering coefficients. In Eqs.~{\bf\ref{clocal}}--{\bf\ref{cglobal}}, and in our subsequent formulas for clustering coefficients, we are of course requiring denominators to be nonzero (as in the monoplex case). In situations in which a denominator vanishes, we set the value of the associated clustering coefficient to $0$.

We can decompose the expression in Eq.~{\bf\ref{cglobal}} in terms of the contributions from cycles that traverse exactly one, two, and three layers (where $m=1, 2, 3$ indicates the number of layers) to give
\begin{align}\label{six}
	t_{*,\micronode} &= t_{*,1,\micronode} \beta^3 + t_{*,2,\micronode} \beta \gamma^2 + t_{*,3,\micronode} \gamma^3\,, \\
	d_{*,\micronode} &= d_{*,1,\micronode} \beta^3 + d_{*,2,\micronode} \beta \gamma^2 + d_{*,3,\micronode} \gamma^3\,, \\
	C_*^{(m)} &= \frac{\sum_{\micronode} t_{*,m,\micronode}}{\sum_{\micronode} d_{*,m,\micronode}}\,.
\end{align}
We can similarly decompose Eqs.~{\bf\ref{clocal}} and~{\bf\ref{suprac}}.
Using the decomposition in Eq.~{\bf\ref{six}} yields an alternative way to average over contributions from the three types of cycles:
\begin{align}\label{ten}
	C_*(\omega_1,\omega_2,\omega_3)=\sum_m^3 \omega_m C_*^{(m)}\,,
\end{align}
where $\vec{\omega}$ is a vector that gives the relative weights of the different contributions. We use the term \emph{layer-decomposed clustering coefficients} for $C_*^{(1)}$, $C_*^{(2)}$, and $C_*^{(3)}$. There are also analogs of Eq.~{\bf\ref{ten}} for the clustering coefficients defined in Eqs.~{\bf\ref{clocal}} and~{\bf\ref{suprac}}. Each of the clustering coefficients in Eqs.~{\bf\ref{clocal}}--{\bf\ref{cglobal}} depends on the values of the parameters $\beta$ and $\gamma$, but the dependence vanishes if $\beta=\gamma$. Unless we explicitly indicate otherwise, we assume in our calculations that $\beta=\gamma$.


\subsection{Clustering Coefficients for Aggregated Networks}
\label{sec:ccagg1}

A common way to study multiplex networks is to aggregate layers to obtain either multi-graphs or weighted networks, where the number of edges or the weight of an edge is the number of different types of edges between a pair of nodes~\cite{Kivela2014Multilayer}. One can then use any of the numerous ways to define clustering coefficients for weighted monoplex networks \cite{Saramaki2007Generalizations,Opsahl2009Clustering} to calculate clustering coefficients for the aggregated network.

One of the weighted clustering coefficients is a special case of our multiplex clustering coefficient (for others, see \ref{sec:weightedcc}).
References~\cite{Zhang2005General,Ahnert2007Ensemble,Grindrod2002Rangedependent} calculated a weighted clustering coefficient as
\begin{align}
	C_{Z,\supranode[1] } &=
\frac{\sum_{\supranode[2]\supranode[3]} W_{\supranode[1]\supranode[2]}W_{\supranode[2]\supranode[3]}W_{\supranode[3]\supranode[1]}}{w_{\max}\sum_{\supranode[2] \neq \supranode[3]} W_{\supranode[1]\supranode[2]}W_{\supranode[1]\supranode[3]}}=\frac{(\mathbf{W}^3)_{\supranode[1]\supranode[1]}}{(\mathbf{W}(w_{\max}\mathbf{F})\mathbf{W})_{\supranode[1]\supranode[1]}}\,, \label{czhang}
\end{align}
where $W_{\supranode[1]\supranode[2]}=\sum_{\micronode[1] \in l(\supranode[1]), \micronode[2] \in l(\supranode[2])} \mathcal{A}_{\micronode[1]\micronode[2]}$ is an element of the weighted adjacency matrix $\mathbf{W}$. The elements of $\mathbf{W}$ are the weights of the edges, the quantity $w_{\max}=\max_{\supranode[1],\supranode[2]}{W_{\supranode[1]\supranode[2]}}$ is the maximum weight in $\mathbf{W}$, and $\mathbf{F}$ is the adjacency matrix of the complete unweighted graph. We can define the global version $C_{Z}$ of $C_{Z,\supranode[1]}$ by summing over all of the nodes in the numerator and the denominator of Eq.~{\bf\ref{czhang}} (analogously to Eq.~{\bf\ref{cglobal}}).

For node-aligned multiplex networks, the clustering coefficients $C_{Z,\supranode[1]}$ and $C_{Z}$ are related to our multiplex clustering coefficients $C_{M,\supranode[1]}$ and $C_{M}$. Letting $\beta=\gamma=1$ and summing over all layers yields $\sum_{i\in l(\supranode[1])} ((\mathcal{A\widehat{C}})^3)_{ii} = (\mathbf{W}^3)_{\supranode[1]\supranode[1]}$. That is, in this special case, the weighted clustering coefficients $C_{Z,\supranode[1]}$ and $C_{Z}$ are equivalent to the corresponding multiplex clustering coefficients $C_{M,\supranode[1]}$ and $C_{M}$. In particular, $C_{M,\supranode[1]}(\beta=\gamma)=\frac{w_{\max}}{b} C_{Z,\supranode[1]}$ and $C_{M}(\beta=\gamma)=\frac{w_{\max}}{b} C_{Z}$. We need the term ${w_{\max}}/{b}$ to match the normalizations because aggregation removes the information about the number of layers $b$, so the normalization must be based on the maximum weight instead of the number of layers. 
That is, a step in the complete weighted network is described by using $w_{\max}\mathbf{F}$ in Eq.~{\bf\ref{czhang}} instead of using $b\mathbf{F}$ .

Note that this relationship between our multiplex clustering coefficient and the weighted clustering coefficient in Eq.~{\bf\ref{czhang}} is only true for node-aligned multiplex networks. If some nodes are not shared among all layers, then the normalization of our multiplex clustering coefficient depends on how many nodes are present in the local neighborhood of the focal node. This contrasts with the ``global'' normalization by $w_{\max}$ used by the weighted clustering coefficient in Eq.~{\bf\ref{czhang}}.


\subsection{Clustering Coefficients in Erd\H{o}s-R\'{e}nyi (ER) networks}
\label{sec:ercc}

Almost all real networks contain some amount of transitivity, and it is often desirable to know if a network contains more transitivity than would be expected by chance. In order to examine this question, one typically compares clustering-coefficient values of a network to what would be expected from some random network that acts as a null model.  The simplest random network to use is an Erd\H{o}s-R\'{e}nyi (ER) network. In this section, we give formulas for expected clustering coefficients in node-aligned multiplex networks in which each intra-layer network is an ER network that is created independently of other intra-layer networks and the inter-layer connections are created as described in Section \ref{sec:mathematical}.

The expected value of the local clustering coefficient in an unweighted monoplex ER network is equal to the probability $p$ of an edge to exist. That is, the density of the neighborhood of a node, measured by the local clustering coefficient, has the same expectation as the density of the entire network for an ensemble of ER networks. In multiplex networks with ER intra-layer graphs with connection probabilities $p_\alpha$, the same result holds only when all of the layers are statistically identical (i.e., $p_\alpha=p$ for all $\alpha$). Note that this is true even if the network is not node-aligned.
However, heterogeneity among layers complicates the behavior of clustering coefficients. If the layers have different connection probabilities, then the expected value of the mean clustering coefficient is a nontrivial function of the connection probabilities. In particular, it is not always equal to the mean of the connection probabilities. For example, the formulas for the expected global layer-decomposed clustering coefficients are
\begin{align}
	\ensavg{C_{M}^{(1)}} &= \frac{\sum_{\alpha}p_{\alpha}^3}{\sum_{\alpha}p_{\alpha}^2} \equiv \frac{\avg{p^3}}{\avg{p^2}}\,, \label{12} \\
	\ensavg{C_{M}^{(2)}} &= \frac{3\sum_{\alpha\neq\kappa}p_{\alpha}p_{\kappa}^2}{(b-1)\sum_{\alpha}p_{\alpha}^2+2\sum_{\alpha\neq\kappa}p_{\alpha}p_{\kappa}} \,, \label{13}\\
	\ensavg{C_{M}^{(3)}} &= \frac{\sum_{\alpha\neq\kappa,\kappa\neq\mu,\mu\neq\alpha} p_{\alpha}p_{\kappa}p_{\mu}}{(b-2)\sum_{\alpha\neq\kappa}p_{\alpha}p_{\kappa}}\,. \label{14}
\end{align}
See \ref{sec:ercc2} for analogous formulas for the local multiplex clustering coefficients and for the numerical validation of our theoretical results.


\section{Results and Discussions}

We investigate transitivity in empirical multiplex networks by calculating clustering coefficients. In Table~\ref{table:ccvalues}, we give the values of layer-decomposed global clustering coefficients for multiplex networks (four social networks and two transportation networks) constructed from real data. Note that the two transportation networks have different numbers of nodes in different layers (i.e., they are not node-aligned  \cite{Kivela2014Multilayer}). 
To help give context to the values, the table also includes the clustering-coefficient values that we obtain for ER networks with matching edge densities in each layer. See \ref{sec:interlayershuffle} for a similar table that uses an alternative null model in which we shuffle the inter-layer connections.

As we will now discuss, multiplex clustering coefficients give insights that are impossible to infer by calculating weighted clustering coefficients for aggregated networks or by calculating clustering coefficients separately for each layer of a multiplex network.

\begin{sidewaystable}
\vspace{15cm}
\begin{tabular*}{\hsize}{@{\extracolsep{\fill}}lccccccc}
\multicolumn{2}{l}{}  &Tailor Shop&Management&Families&Bank&Tube&Airline\\
\hline
\multirow{2}{*}{$C_{M}$}&orig.&0.319**&0.206**&0.223'&0.293**&0.056&0.101** \\
                       &ER&$0.186\pm 0.003$&$0.124\pm 0.001$&$0.138\pm 0.035$&$0.195\pm 0.009$&$0.053\pm 0.011$&$0.038\pm 0.000$ \\
\hline
\multirow{2}{*}{$C_{M}^{(1)}$}&orig.&0.406**&0.436**&0.289'&0.537**&0.013''&0.100** \\
                            &ER&$0.244\pm 0.010$&$0.196\pm 0.015$&$0.135\pm 0.066$&$0.227\pm 0.038$&$0.053\pm 0.013$&$0.064\pm 0.001$ \\
\hline
\multirow{2}{*}{$C_{M}^{(2)}$}&orig.&0.327**&0.273**&0.198&0.349**&0.043*&0.150** \\
                            &ER&$0.191\pm 0.004$&$0.147\pm 0.002$&$0.138\pm 0.040$&$0.203\pm 0.011$&$0.053\pm 0.020$&$0.041\pm 0.000$ \\
\hline
\multirow{2}{*}{$C_{M}^{(3)}$}&orig.&0.288**         &0.192**         &-&0.227**&0.314**&0.086** \\
                            &ER   &$0.165\pm 0.004$&$0.120\pm 0.001$&-&$0.186\pm 0.010$&$0.051\pm 0.043$&$0.037\pm 0.000$ \\
\end{tabular*}
\caption{Clustering coefficients $C_{M}$, $C_{M}^{(1)}$, $C_{M}^{(2)}$, and $C_{M}^{(3)}$ that correspond, respectively, to the global, one-layer, two-layer, and three-layer clustering coefficients for various multiplex networks. ``Tailor Shop'': Kapferer tailor-shop network ($n=39$, $b=4$)~\cite{Kapferer1972Strategy}. ``Management'': Krackhardt office cognitive social structure ($n=21$, $b=21$)~\cite{Krackhardt1987Cognitive}. ``Families'': Padgett Florentine families social network ($n=16$, $b=2$)~\cite{Breiger1986Cumulated}. ``Bank'': Roethlisberger and Dickson bank wiring-room social network ($n=14$, $b=6$)~\cite{Roethlisberger1939Management}. ``Tube'': The London Underground (i.e., ``The Tube'') transportation network ($n=314$, $b=14$)~\cite{Rombach2012Core}. ``Airline'': Network of flights between cities, in which each layer corresponds to a single airline ($n=3108$, $b=530$)~\cite{airlinedata}.
The rows labeled ``orig.'' give the clustering coefficients for the original networks, and the rows labeled ``ER'' give the expected value and the standard deviation of the clustering coefficient in an ER random network with exactly as many edges in each layer as in the original network. For the original values, we perform a two-tailed Z-test to examine whether the observed clustering coefficients could have been produced by the ER networks. We designate the p-values as follows: *: $p<0.05$, **: $p<0.01$ for Bonferroni-corrected tests with 24 hypothesis; ': $p<0.05$, '': $p<0.01$ for uncorrected tests. We do not use any symbols for values that are not significant. We symmetrize directed networks by considering two nodes to be adjacent if there is at least one edge between them. The social networks in this table are node-aligned multiplex graphs, but the transportation networks are not node-aligned. We report values that are means over different numbers of realizations: $1.5 \times 10^5$ for Tailor Shop, $1.5 \times 10^3$ for Airline, $1.5 \times 10^4$ for Management, $1.5 \times 10^5$ for Families, $1.5 \times 10^4$ for Tube, and $1.5 \times 10^5$ for Bank.}
\label{table:ccvalues}
\end{sidewaystable}

For each social network in Table~\ref{table:ccvalues}, note that $C_{M}<C_M^{(1)}$ and $C_M^{(1)} > C_M^{(2)} > C_M^{(3)}$.  Consequently, the primary contribution to the triadic structure of these multiplex networks arises from 3-cycles that stay within a given layer. To check that the ordering of the different clustering coefficients is not an artifact of the heterogeneity of densities of the different layers, we also calculate the expected values of the clustering coefficients in ER networks with identical edge densities to the data. We observe that all clustering coefficients exhibit larger inter-layer transitivities than would be expected in corresponding ER networks with identical edge densities, although the same ordering relationship (i.e. $C_M^{(1)} > C_M^{(2)} > C_M^{(3)}$ ) holds. 
From our results in Table \ref{table:ccvalues}, it seems that triadic-closure mechanisms in social networks cannot be considered purely at the aggregated network level; these mechanisms appear to be more effective inside of layers than between layers.
For example, if there is a connection between individuals $u$ and $v$ and also a connection between $v$ and $w$ in the same layer, then it is more likely that $u$ and $w$ ``meet'' in the same layer than in some other layer.

The transportation networks that we examine exhibit the opposite pattern from the social networks. For example, for the London Underground (``Tube'') network, in which each layer corresponds to a line, we observe that $C_M^{(3)} > C_M^{(2)} > C_M^{(1)}$. This reflects the fact that single lines in the Tube are designed to avoid redundant connections.
A single-layer triangle would require a line to make a loop among 3 stations. Two-layer triangles, which are a bit more frequent than single-layer ones, entail that two lines run in almost parallel directions and that one line jumps over a single station. For 3-layer triangles, the geographical constraints do not matter because one can construct a triangle with three straight lines.

\begin{figure}[t]
\begin{center}
\includegraphics[width=8.7cm]{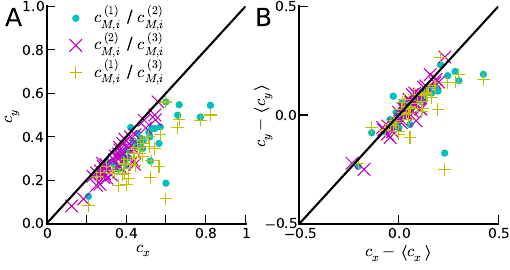}
\caption{Comparison of different local clustering coefficients in the Kapferer tailor-shop network. Each point corresponds to a node. (A) The raw values of the clustering coefficients. (B) The value of the clustering coefficients minus the expected value of the clustering coefficient for the corresponding node from a mean over 1000 realizations of a configuration model with the same degree sequence in each layer as in the original network. In a realization of the multiplex configuration model, each intra-layer network is an independent realization of the monoplex configuration model.
}
\label{fig:emprical_comparisons}
\end{center}
\end{figure}

We also analyze the local triadic closure of the Kapferer tailor-shop social network by examining the local clustering-coefficient values. In Fig.~\ref{fig:emprical_comparisons}A, we show a comparison of the layer-decomposed local clustering coefficients (also see Fig.~6a of \cite{Battiston2014Metrics}). Observe that the condition $c_{M,\micronode}^{(1)} > c_{M,\micronode}^{(2)} > c_{M,\micronode}^{(3)}$ holds for most of the nodes. In Fig.~\ref{fig:emprical_comparisons}B, we subtract the expected values of the clustering coefficients of nodes in a network generated with the configuration model\footnote{We use the configuration model instead of an ER network as a null model because the local clustering-coefficient values are typically correlated with node degree in monoplex networks \cite{newman2010}, and an ER-network null model does not preserve degree sequence.} from the corresponding clustering-coefficient values observed in the data to discern whether we should also expect to observe the relative order of the local clustering coefficients in an associated random network (with the same layer densities and degree sequences as the data). Similar to our results for global clustering coefficients, we see that taking a null model into account lessens---but does not remove---the difference between the coefficients that count different numbers of layers.

\begin{figure}[t]
\begin{center}
\includegraphics[width=9.1cm]{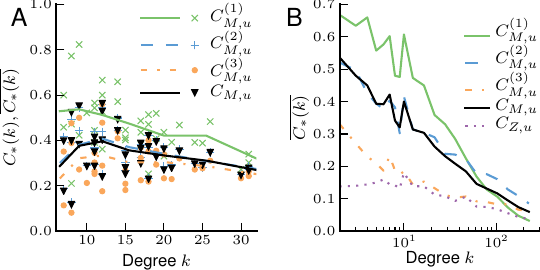}
\caption{Local clustering coefficients versus unweighted degree of the aggregated network for (A) the Kapferer tailor-shop network and (B) the airline network. The curves give the mean values of the clustering coefficients for a degree range (i.e., we bin similar degrees). Note that the horizontal axis in panel (B) is on a logarithmic scale.
}
\end{center}
\label{fig:empirical_degree}
\end{figure}

We investigate the dependence of local triadic structure on degree for one social network and one transportation network. In Fig.~\ref{fig:empirical_degree}A, we show how the different multiplex clustering coefficients depend on the unweighted degrees of the nodes in the aggregated network for the Kapferer tailor shop. Note that the relative ordering of the values of the mean clustering coefficient does not depend on degree. In Fig.~\ref{fig:empirical_degree}B, we illustrate that the aggregated network for the airline transportation network exhibits a non-constant difference between the curves of $C_{M,\supranode}$ and the weighted clustering coefficient $C_{Z,\supranode}$. Using a global normalization (see the discussion in Section \ref{sec:ccagg1}) reduces the clustering-coefficient values for the small airports much more than it does for the large airports. This, in turn, introduces a bias.

The airline network is organized differently from the London Tube network. When comparing these networks, note that each layer in the former encompasses flights from a single airline. For the airline network (see Fig.~\ref{fig:empirical_degree}B), we observe that the two-layer local clustering coefficient is larger than the single-layer one for hubs (i.e., high-degree nodes), but it is smaller for small airports (i.e., low-degree nodes). However, the global clustering coefficient counts the total number of 3-cycles and connected triplets, and it thus gives more weight to high-degree nodes than to low-degree nodes. We thus find that the global clustering coefficients for the airline network satisfy $C_M^{(2)} > C_M^{(1)} > C_M^{(3)}$. The intra-airline clustering coefficients have small values, presumably because it is not in the interest of an airline to introduce new flights between two airports that can already be reached by two flights via the same airline through some major airport. The two-layer cycles correspond to cases in which an airline has a connection from an airport to two other airports and a second airline has a direct connection between those latter two airports. Completing a three-layer cycle requires using three distinct airlines, and this type of congregation of airlines to the same area is not frequent in the data. Three-layer cycles are more common than single-layer cycles only for a few of the largest airports.

\section{Conclusions}

We derived measurements of transitivity for multiplex networks by developing multiplex generalizations of triadic relationships and clustering coefficients.  By using examples from empirical data in diverse settings, we showed that different notions of multiplex transitivity are important in different situations.  For example, the balance between intra-layer versus inter-layer clustering is different in social networks versus transportation networks (and even in different types of networks within each category, as we illustrated explicitly for transportation networks), reflecting the fact that multilayer transitivity can arise from different mechanisms.  Such differences are rooted in the new degrees of freedom that arise from inter-layer connections and are invisible to calculations of clustering coefficients on single-layer networks obtained via aggregation.  In other words, transitivity is inherently a multilayer phenomenon: all of these diverse flavors of transitivity reduce to the same description when one throws away the multilayer information. Generalizing clustering coefficients for multiplex networks makes it possible to explore such phenomena and to gain deeper insights into different types of transitivity in networks. The existence of multiple types of transitivity also has important implications for multiplex network motifs and multiplex community structure.  In particular, our work on multiplex clustering coefficients demonstrates that the definition of any clustering notion for multiplex networks needs to be able to consider diverse forms of transitivity.


\section{Acknowledgements}

All authors were supported by the European Commission FET-Proactive project PLEXMATH (Grant No. 317614). AA also acknowledges financial support from the ICREA Academia, Generalitat de Catalunya (2009-SGR-838), and the James S.\ McDonnell Foundation; and SG and AA were supported by FIS2012-38266. YM was also supported by MINECO through Grants FIS2011-25167 and by DGA (Spain). MAP acknowledges a grant (EP/J001759/1) from the EPSRC. We thank David Krackhardt and the anonymous referees for useful comments.

\appendix










\clearpage




\section{Weighted Clustering Coefficients}
\label{sec:weightedcc}

There are two primary weighted clustering coefficients for monoplex networks that provide alternatives to the one that we discussed in the main text~\cite{Onnela2005Intensity,Barrat2004Architecture}. They are
\begin{align}
	C_{O,\supranode[1]} &= \frac{1}{w_{\max}k_{\supranode[1]}(k_{\supranode[1]}-1)} \sum_{\supranode[2],\supranode[3]} (W_{\supranode[1]\supranode[2]}W_{\supranode[1]\supranode[3]}W_{\supranode[2]\supranode[3]})^{1/3}\,,\label{connela}
\\
	C_{Ba,\supranode[1]} &=
\frac{1}{s_{\supranode[1]}(k_{\supranode[1]}-1)}\sum_{\supranode[2],\supranode[3]}\frac{(W_{\supranode[1]\supranode[2]}+W_{\supranode[1]\supranode[3]})}{2}
A_{\supranode[1]\supranode[2]}A_{\supranode[1]\supranode[3]}A_{\supranode[2]\supranode[3]}\,,\label{cbarrat}
\end{align}
where ${\bf A}$ is the unweighted adjacency matrix associated with the weighted adjacency matrix ${\bf W}$, the degree of node $\supranode[1]$ is $k_{\supranode[1]}=\sum_{\supranode[2]}A_{\supranode[1]\supranode[2]}$, the strength of $\supranode[1]$ is $s_{\supranode[1]}=\sum_{\supranode[2]}W_{\supranode[1]\supranode[2]}$, and the quantity $w_{\max}=\max_{\supranode[1],\supranode[2]}{W_{\supranode[1]\supranode[2]}}$ is the maximum weight in ${\bf W}$. When using Eqs.~{\bf\ref{connela}} and {\bf\ref{cbarrat}}, one also has $C_{O,\supranode[1]} = 0$ and $C_{Ba,\supranode[1]} = 0$ for nodes of degree $k_{\supranode[1]} = 0$ and $k_{\supranode[1]} = 1$.


\section{Multiplex Clustering Coefficients in the Literature}
\label{sec:mplexcc}

Let $\mathbf{A}^{(\alpha)}$ denote the intra-layer adjacency matrix for layer $\alpha$.
For a weighted multiplex network, we use $\mathbf{W}^{(\alpha)}$ denote the intra-layer weight matrix (i.e., the weighted intra-layer adjacency matrix) for layer $\alpha$. We use $\mathbf{W}$ to denote the weight matrix of the aggregated network. (See Section~\ref{sec:ccagg1} in the main text.)
 The clustering coefficient that was defined in \cite{Barrett2012Taking} for node-aligned multiplex networks is
\begin{align}
	C_{Be,\supranode[1]}=\frac{\sum_{\supranode[2],\supranode[3]} \sum_\alpha  A_{\supranode[1]\supranode[2]}^{(\alpha)} \sum_\kappa A_{\supranode[1]\supranode[3]}^{(\kappa)} \sum_\mu A_{\supranode[2]\supranode[3]}^{(\mu)}  } {\sum_{\supranode[2],\supranode[3]} \sum_\kappa A_{\supranode[1]\supranode[2]}^{(\kappa)} \sum_\alpha \max(A_{\supranode[1]\supranode[3]}^{(\alpha)},A_{\supranode[2]\supranode[3]}^{(\alpha)}) }\,,
\end{align}
which can be expressed in terms of the aggregated network as
\begin{align}
	C_{Be,\supranode[1]} =\frac{\sum_{\supranode[2],\supranode[3]} W_{\supranode[1]\supranode[2]}W_{\supranode[1]\supranode[3]}W_{\supranode[2]\supranode[3]}} {\sum_{\supranode[2],\supranode[3]} W_{\supranode[1]\supranode[2]} \sum_\alpha \max\left(A_{\supranode[1]\supranode[3]}^{(\alpha)},A_{\supranode[2]\supranode[3]}^{(\alpha)}\right) }\,.
\label{cbarrettsimple}
\end{align}
The numerator of Eq.~{\bf\ref{cbarrettsimple}} is the same as the numerator of the weighted clustering coefficient $C_{Z,\supranode[1]}$, but the denominator is different. Because of the denominator in Eq.~{\bf\ref{cbarrettsimple}}, the values of the clustering coefficient $C_{Be,\supranode[1]}$ do not have to lie in the interval $[0,1]$.  For example, $C_{Be,\supranode[1]} = (n-2)b/n$ for a complete multiplex network (where $n$ is the number of nodes in the multiplex network), so $C_{Be,\supranode[1]} > 1$ when $b > \frac{n}{n-2}$.

References~\cite{Brodka2010Method,Brodka2012Analysis} defined a family of local clustering coefficients for directed and weighted multiplex networks:
\begin{align}\label{this}
	C_{Br,u,t} = \frac{\sum_{\alpha \in L} \sum_{v,w \in N(u,t)} ( W_{wv}^{(\alpha)} + W_{vw}^{(\alpha)})}{2 |N(u,t)| b}\,,
\end{align}
where $N(u,t) = \{ v : | \{\alpha : A_{uv}^{(\alpha)}=1 \quad \mathrm{and} \quad A_{vu}^{(\alpha)}=1 \}| \geq t \}$, $t$ is a threshold, and we recall that $L = \{1, \ldots, b\}$ is the set of layers.
The clustering coefficient {\bf\ref{this}} does not yield the ordinary monoplex local clustering coefficient for unweighted (i.e., networks with binary weights) and undirected networks when it is calculated for the special case of a monoplex network (i.e., a multiplex network with $b = 1$ layer). Furthermore, its values are not normalized to lie between $0$ and $1$. For example, consider a complete multiplex network with $n$ nodes and an arbitrary number of layers. In this case, the clustering coefficient {\bf \ref{this}} takes the value of $n-2$ for each node.
If a multiplex network is undirected (and unweighted), then $C_{Br,u,t}$ can always be calculated when one is only given an aggregated network and the total number of layers in the multiplex network. As an example, for the threshold value $t=1$, one obtains
\begin{align}
	C_{Br,\supranode[1],1}=\frac{1}{k_{\supranode[1]}b}\sum_{\supranode[2],\supranode[3]}\frac{W_{\supranode[2]\supranode[3]}}{2}A_{\supranode[1]\supranode[2]}A_{\supranode[1]\supranode[3]}A_{\supranode[2]\supranode[3]}\,,
\label{brsimple}
\end{align}
where ${\bf A}$ is the binary adjacency matrix corresponding to the weighted adjacency matrix ${\bf W}$ and $k_{\supranode[1]}=\sum_{\supranode[2]}A_{\supranode[1]\supranode[2]}$ is the degree of node $\supranode[1]$.

Reference~\cite{Criado2011Mathematical} defined a clustering coefficient for multiplex networks that are not necessarily node-aligned as
\begin{align}
	C_{Cr,u} = \frac{2 \sum_{\alpha \in L} | \overline{E}_\alpha(u) | }{\sum_\alpha  |\Gamma_\alpha (u)| (|\Gamma_\alpha (u)| -1)}\,,
\end{align}
where $L = \{1,\ldots,b\}$ is again the set of layers, $\Gamma_\alpha (u)=\Gamma (u) \cap V_\alpha$, the quantity $\Gamma (u)$ is the set of neighbors of node $u$ in the aggregated network, $V_\alpha$ is the set of nodes in layer $\alpha$, and $\overline{E}_\alpha(u)$ is the set of edges in the subgraph induced by $\Gamma_\alpha (u)$ in the aggregated network. For a node-aligned multiplex network, $V_\alpha=V$ and  $\Gamma_\alpha (u)=\Gamma (u)$, so one can write
%
\begin{align}
	C_{Cr,\supranode[1]}=\frac{\sum_{\supranode[2]\supranode[3]}A_{\supranode[1]\supranode[2]}W_{\supranode[2]\supranode[3]}A_{\supranode[3]\supranode[1]}}{b \sum_{\supranode[2] \neq \supranode[3]}A_{\supranode[1]\supranode[2]}A_{\supranode[3]\supranode[1]}}\,,
\end{align}
which is a local clustering coefficient for the aggregated network.

Battiston et al.~\cite{Battiston2014Metrics} defined two versions of clustering coefficients for node-aligned multiplex networks:
\begin{align}
	C_{Bat1,u} &= \frac{\sum_{\alpha} \sum_{\kappa \neq \alpha } \sum_{v \neq u, w \neq u} A_{uv}^{(\alpha)} A_{vw}^{(\kappa)} A_{wu}^{(\alpha)}}{(b-1)\sum_{\alpha}\sum_{v \neq u, w \neq u} A_{uv}^{(\alpha)} A_{wu}^{(\alpha)}} \,, \label{battiston1}\\
	C_{Bat2,u} &= \frac{\sum_{\alpha} \sum_{\kappa \neq \alpha } \sum_{\mu \neq \alpha, \kappa } \sum_{v \neq u, w \neq u} A_{uv}^{(\alpha)} A_{vw}^{(\mu)} A_{wu}^{(\kappa)}}{(b-2)\sum_{\alpha} \sum_{\kappa \neq \alpha }\sum_{v \neq u, w \neq u} A_{uv}^{(\alpha)} A_{wu}^{(\kappa)}} \,.
\label{battiston2}
\end{align}
The first definition, $C_{Bat1,u}$, counts the number of ${\cal ACACA}$-type elementary cycles; and the second definition, $C_{Bat2,u}$, counts the 3-layer elementary cycles ${\cal ACACAC}$. In both of these definitions, the sums in the denominators allow terms in which $v=w$, so a complete multiplex network has a local clustering coefficient of $(n-1)/(n-2)$ for every node.

Reference~\cite{tensorial13} proposed definitions for global clustering coefficients using a tensorial formalism for multilayer networks; when representing a multiplex network as a third-order tensor, the formulas in \cite{tensorial13} reduce to the clustering coefficients that we propose in the present article. (See Eq.~{\bf\ref{cglobal}} of the main text.)
Parshani et al. \cite{Parshani2010Intersimilarity} defined an ``inter-clustering coefficient'' for two-layer interdependent networks that can be interpreted as multiplex networks \cite{Son2011Percolation,Son2012Percolation,Baxter2012Avalanche,Kivela2014Multilayer}. Their definition is similar to edge ``overlap'' \cite{Battiston2014Metrics}; in our framework, it corresponds to counting 2-cycles of type $({\cal AC})^2$. A few other scholars \cite{Donges2011Investigating,Podobnik2012Preferential} have also defined generalizations of clustering coefficients for multilayer networks that cannot be interpreted as multiplex networks \cite{Kivela2014Multilayer}.

In Table~\ref{table:cc_properties}, we show a summary of the properties satisfied by several different (local and global) multiplex clustering coefficients. In particular, we check the following properties. (1) The value of the clustering coefficient reduces to the values of the associated monoplex clustering coefficient for a single-layer network. (2) The value of the clustering coefficient is normalized so that it takes values that are less than or equal to 1. (All of the clustering coefficients are nonnegative.) (3) The clustering coefficient has a value of $p$ in a large (i.e., when the number of nodes $n \rightarrow \infty$) node-aligned multiplex network in which each layer is an independent ER network with an edge probability of $p$ in each layer. (4) Suppose that we construct a multiplex network by replicating the same given monoplex network in each layer. We indicate whether the clustering coefficient for the multiplex network has the same value as for the monoplex network. (5) There exists a version of the clustering coefficient that is defined for each node-layer pair separately. (6) The clustering coefficient is defined for multiplex networks that are not node-aligned.

\begin{table*}[t]
\caption{Summary of the properties of the different multiplex clustering coefficients. The notation $C_{*(,\supranode[1])}$ means that the property holds for both the global version and the local version of the associated clustering coefficient (C.C.).}
\begin{adjustwidth}{-.4in}{0in}  
\begin{center}
\begin{tabular}{@{\vrule height 10.5pt depth4pt  width0pt}lccccccccc}
  \\
  Property                          & $C_{M(,\supranode)}$ &  $C_{Be,\supranode}$ & $C_{Z(,\supranode)}$ & $C_{Ba,\supranode}$ & $C_{O,\supranode}$ & $C_{Br,\supranode}$ & $C_{Cr,\supranode}$ & $C_{Bat(1,2),\supranode}$\\
  \hline
  (1) Reduces to monoplex C.C.        & \checkmark       &  \checkmark       & \checkmark       & \checkmark      & \checkmark     &                 & \checkmark      & \\
  (2) $C_{*} \leq 1$                  & \checkmark       &                   & \checkmark       &\checkmark       & \checkmark     &                 & \checkmark      & \checkmark \\
  (3) $C_{*} = p$ in multiplex ER graph     & \checkmark       &                   &                  &                 &                &                 & \checkmark      &\\
  (4) Monoplex C.C. for copied layers & \checkmark       &                   & \checkmark       &\checkmark       & \checkmark     &                 & \checkmark      &\\
  (5) Defined for node-layer pairs     & \checkmark       &                   &                  &                 &                &                 &                 &\\
  (6) Defined for non-node-aligned      & \checkmark       &                   &                  &                 &                &                 & \checkmark      &\\
\end{tabular}
\label{table:cc_properties}
\end{center}
\end{adjustwidth}
\end{table*}


\section{Other Possible Definitions of Cycles}
\label{sec:othercycles}

There are many possible ways to define cycles in multiplex networks. If we relax the condition of disallowing two consecutive inter-layer steps, then we can write
\begin{align}
	t_{SM,i} &= [(\mathcal{\widehat{C}A\widehat{C}})^3]_{ii}\,, \label{tsaw} \\
	t_{SM',i} &= [(\mathcal{\widehat{C}^\prime A} +\mathcal{A\widehat{C}^\prime})^3]_{ii}\,, \label{tsaw2}
\end{align}
where $\mathcal{\widehat{C}}^\prime=\frac{1}{2}\beta \mathcal{I}+\gamma \mathcal{C}$. 
(Our discussion in Appendix D is helpful for understanding the factor of 1/2.) Unlike 
the matrix $\mathcal{A\widehat{C}}$ in Eq.~{\bf\ref{tawsimple}}
in the main text, the matrices $\mathcal{\widehat{C}A\widehat{C}}$ and $\mathcal{\widehat{C}^\prime A} +\mathcal{A\widehat{C}^\prime}$ are symmetric. We can thus interpret them as weighted adjacency matrices of symmetric supra-graphs, and we can then calculate cycles and clustering coefficients in these supra-graphs (see \ref{sec:auxiliary}).

It is sometimes desirable to forbid the option of staying inside of a layer in the first step of the second term of Eq.~{\bf\ref{tsaw2}}.  In this case, one can write
\begin{align}
	t_{M^\prime,i}=[(\mathcal{A\widehat{C}})^3 + \gamma {\cal CA}(\mathcal{\widehat{C}A})^2]_{ii}\,. \label{tmoreno}
\end{align}
With this restriction, cycles that traverse two edges of the focal node $i$ are only calculated two times instead of four times. 
 In this case, we simplify Eq.~{\bf\ref{tmoreno}} to obtain
\begin{align}
	t_{M^\prime,i}=[2(\mathcal{A\widehat{C}})^2\mathcal{A\widehat{C}^\prime}]_{ii}\,, \label{tmorenosimple}
\end{align}
which is similar to Eq.~{\bf\ref{tawsimple}} in the main text. In Table~\ref{table:cc_values2}, we show the values of the clustering coefficients that we calculate using this last definition of cycle for the empirical networks that we studied in the main text.

\begin{table*}
  \caption{Clustering coefficients (rows) for the same empirical networks (columns) from Table~1 in the main text. For the Tube and the Airline networks, we only calculate clustering coefficients for non-node-aligned networks.
For local clustering coefficients, we average over all nodes to obtain $\avg{C_{*,\supranode[1]}}=\frac{1}{n}\sum_{\supranode[1]}C_{*,\supranode[1]}$.
  }
  \begin{center}
    \begin{tabular}{lcccccc}
      CC &Families&Bank&Tailor Shop&Management&Tube&Airline\\
      \hline
      $C_{M^\prime}$&0.218&0.289&0.320&0.206&0.070& 0.102\\
      $C_{M^\prime}^{(1)}$&0.289&0.537&0.406&0.436&0.013&0.100 \\
      $C_{M^\prime}^{(2)}$&0.202&0.368&0.338&0.297&0.041&0.173 \\
      $C_{M^\prime}^{(3)}$& -   &0.227&0.288&0.192&0.314& 0.086\\
      $C_{M^\prime}(\frac{1}{3},\frac{1}{3},\frac{1}{3})$&0.164&0.377&0.344&0.309&0.123&0.120 \\
      $\avg{C_{Cr,\supranode[1]}}$ &0.342 & 0.254 & 0.308 & 0.150 & 0.038 & 0.329 \\
      $\avg{C_{Ba,\supranode[1]}}$ & 0.195 & 0.811 & 0.612 & 2.019 & - & - \\
      $\avg{C_{Br,u}}$ &0.674 & 1.761 & 4.289 & 1.636 & - & - \\
      $\avg{C_{O,\supranode[1]}}$&0.303&0.268&0.260&0.133& - & - \\
      $\avg{C_{Be,\supranode[1]}}$&0.486&0.775&0.629&0.715& - & - \\
      $\avg{C_{Bat1,\supranode[1]}}$&0.159 & 0.199 & 0.271 & 0.169 & - & - \\
      $\avg{C_{Bat2,\supranode[1]}}$& - &  0.190 & 0.282 & 0.179  & - & - \\
    \end{tabular}
  \end{center}
  \label{table:cc_values2}
\end{table*}


\section{Defining Multiplex Clustering Coefficients Using Auxiliary Networks}
\label{sec:auxiliary}

An elegant way to generalize clustering coefficients for multiplex networks is to define a new (possibly weighted) auxiliary supra-graph $G_M$ so that one can define cycles of interest as weighted 3-cycles in $G_M$.
Once we have a function that produces the auxiliary supra-adjacency matrix $\mathcal{M}=\mathcal{M}(\mathcal{A},\mathcal{C})$, we can define the auxiliary complete supra-adjacency matrix $\mathcal{M}^F=\mathcal{M}(\mathcal{F},\mathcal{C})$. One can then define a local clustering coefficient for node-layer pair $i$ with the formula
\begin{align}
	c_{\micronode[1]}=\frac{(\mathcal{M}^3)_{\micronode[1]\micronode[1]}}{(\mathcal{MM}^F\mathcal{M})_{\micronode[1]\micronode[1]}}\,.
\end{align}
As with a monoplex network, the denominator written in terms of the complete matrix $\mathcal{M}^F$ is equivalent to the usual one written in terms of connectivity. We thereby consider the connectivity of a node in the supra-graph induced by
the matrix $\mathcal{M}$. We refer to the matrix $\mathcal{M}$ as the \emph{multiplex walk matrix} because it encodes the permissible steps in a multiplex network. When $\mathcal{M}$ is equal to $\mathcal{A\widehat{C}}$ or to $\mathcal{\widehat{C}A}$, the induced supra-graph is directed, so one needs to distinguish between in-degrees and out-degrees.

A key advantage of defining clustering coefficients using an auxiliary supra-graph is that one can then use it to calculate other diagnostics (e.g., degree or strength) for nodes. One can thereby investigate correlations between clustering-coefficient values and the size of the multiplex neighborhood of a node.  (The size of the neighborhood is the number of nodes that are reachable in a single step via connections defined by the matrix $\mathcal{M}$.)

We can write the symmetric multiplex walk matrices in Eqs.~{\bf\ref{tsaw}} and~{\bf\ref{tsaw2}} as
\begin{align}
	\mathcal{M}_{SM}=&\mathcal{\widehat{C}A\widehat{C}}\,,\label{saw}\\
	\mathcal{M}_{SM^\prime}=&(\mathcal{\widehat{C}^\prime A} + \mathcal{A\widehat{C}^\prime})\,.\label{sawprime}
\end{align}
To avoid double-counting intra-layer steps in the definition of $\mathcal{M}_{SM^\prime}$, we need to rescale either the intra-layer weight parameter $\beta$ (i.e., we can write $\mathcal{\widehat{C}^\prime}=\beta^\prime \mathcal{I}+\gamma \mathcal{C}=\frac{1}{2}\beta \mathcal{I}+\gamma \mathcal{C}$) or the inter-layer weight parameter $\gamma$ [i.e., we can write $\mathcal{\widehat{C}^\prime}=\beta \mathcal{I}+\gamma^\prime \mathcal{C}=\beta \mathcal{I}+2\gamma \mathcal{C}$ and also define $\mathcal{M}_{SM^\prime}=\frac{1}{2}(\mathcal{A\widehat{C}^\prime}+\mathcal{\widehat{C}^\prime A})$].

Consider a supra-graph induced by a multiplex walk matrix.  The distinction between the matrices $\mathcal{M}_{SM}$ and $\mathcal{M}_{SM^\prime}$ is that $\mathcal{M}_{SM}$ also includes terms of the form $\mathcal{CAC}$ that take into account walks that have an inter-layer step ($\mathcal{C}$) followed by an intra-layer step ($\mathcal{A}$) and then another inter-layer step ($\mathcal{C}$).  Therefore, in the supra-graph induced by $\mathcal{M}_{SM}$, two nodes in the same layer that are not adjacent in that layer are nevertheless adjacent if the same physical nodes are adjacent in another layer.

The matrix $\mathcal{\widehat{C}}$ sums the contributions of all node-layer pairs that correspond to the same physical node when $\beta=\gamma=1$. In other words, if we associate a vector of the canonical basis $e_i$ to each node-layer pair $i$ and let $\Gamma_C((u,\alpha))=\{ (u,\kappa) | \kappa \in L \}$ denote all node-layer pairs that correspond to the same physical node, then
\begin{align}
	\mathcal{\widehat{C}}e_{i} = \sum_{j \in \Gamma_C(i)} e_j
\end{align}
produces a vector whose entries are equal to $1$ for nodes that belong to the basis vector and which are equal to $0$ for nodes that do not belong to that vector.
Consequently, $\mathcal{M}_{SM}$ is related to the weighted adjacency matrix of the aggregated graph for $\beta=\gamma=1$.  To be precise, we obtain the following relation:
\begin{align}
	(\mathcal{\widehat{C}A\widehat{C}})_{ij} = W_{\supranode[1]\supranode[2]}\,, \, \quad \ \text{for any}  \,\, i \in l(\supranode[1]), j \in l(\supranode[2] ) \,.
\end{align}

One can also write the multiplex clustering coefficient induced by Eq.~{\bf\ref{taw}}
in terms of the auxiliary supra-adjacency matrix by considering Eq.~{\bf\ref{tawsimple}},
which is a simplified version of the equation that counts cycles only in one direction. This yields
\begin{align}
	\mathcal{M}_{M}=\sqrt[3]{2}\mathcal{A\widehat{C}}\,.
	\label{eq20}
\end{align}
The matrix $\mathcal{M}_{M}$ is not symmetric, which implies that the associated graph is a directed supra-graph. Nevertheless, the clustering coefficient induced by $\mathcal{M}_{M}$ is the same as that induced by its transpose $\mathcal{M}_{M}^T$ if $\mathcal{A}$ is symmetric.


\section{Expressing Clustering Coefficients Using Elementary 3-Cycles}
\label{sec:elemcycles}

We now give a detailed explanation of the process of decomposing any of our walk-based clustering coefficients into elementary cycles. An \emph{elementary cycle} is a term that consists of products of the matrices $\mathcal{A}$ and $\mathcal{C}$ (i.e., there are no sums) after one expands the expression for a cycle (which is a weighted sum of such terms). Because we are only interested in the diagonal elements of the terms and we consider only undirected intra-layer supra-graphs and coupling supra-graphs, we can transpose the terms and still write them in terms of the matrices $\mathcal{A}$ and $\mathcal{C}$ rather than also using their transposes. There are also multiple ways of writing non-symmetric elementary cycles [e.g., $({\cal AACAC})_{ii}=({\cal CACAA})_{ii}$].

We adopt a convention in which we transpose all elementary cycles so that we select the one in which the first element is $\mathcal{A}$ rather than $\mathcal{C}$ when comparing the two versions of the term from left to right. That is, for two equivalent terms, we choose the one that comes first in alphabetical order.
To calculate the clustering coefficients that we defined in the appendix (see \ref{sec:othercycles} and \ref{sec:auxiliary}), we also need to include elementary cycles that start and end in an inter-layer step. The set of elementary 3-cycles is thus $\mathscr{E} = \{ \mathcal{AAA}$, $\mathcal{AACAC}$, $\mathcal{ACAAC}$, $\mathcal{ACACA}$, $\mathcal{ACACAC}$, $\mathcal{CAAAC}$, $\mathcal{CAACAC}$, $\mathcal{CACACAC} \} $.

\begin{figure*}
 \includegraphics[width=0.95\textwidth]{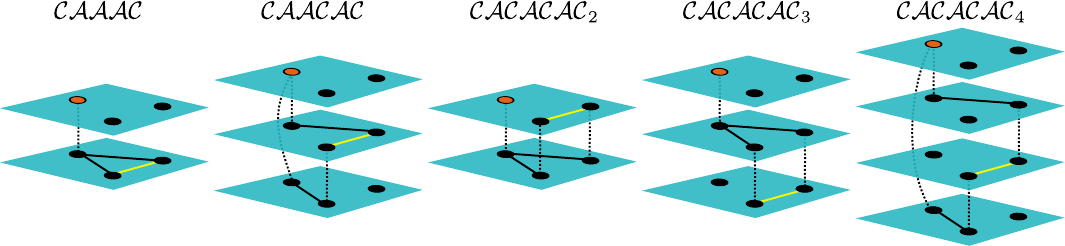}
  \caption{Sketches of elementary cycles for which both the first and the last step are allowed to be an inter-layer step. These elementary cycles are $\mathcal{CAAAC}$, $\mathcal{CAACAC}$, and $\mathcal{CACACAC}$. The orange node is the starting point of the cycle. The intra-layer edges are the solid lines, and the intra-layer edges are the dotted curves. In each case, the yellow line represents the second intra-layer step. Note that the elementary cycle $\mathcal{CACACAC}$ includes three ``degenerate'' versions in which the 3-cycle returns to a previously-visited layer. The subscripts in the names of the degenerate cycles indicate the number of layers that are used in each cycle.}
  \label{fig:elementary_cycles_others}
\end{figure*}

We now write our clustering coefficients using elementary 3-cycles. We obtain the normalization formulas by using the elementary 3-cycles and then replacing the second $\mathcal{A}$ term with $\mathcal{F}$. This yields a standard form for any of our multiplex clustering coefficients. For example,
\begin{equation}
	c_{*,i} = \frac{t_{*,i}}{d_{*,i}}\,,
\end{equation}
where
\begin{align}
\nonumber
	t_{*,i} &=[ w_{\mathcal{AAA}} \mathcal{AAA} + w_{\mathcal{AACAC}} \mathcal{AACAC} + w_{\mathcal{ACAAC}} \mathcal{ACAAC} \\ \nonumber
   &\qquad  +w_{\mathcal{ACACA}} \mathcal{ACACA} + w_{\mathcal{ACACAC}} \mathcal{ACACAC}  \\ \nonumber
   &\qquad  +w_{\mathcal{CAAAC}} \mathcal{CAAAC} +  w_{\mathcal{CAACAC}} \mathcal{CAACAC}  \\
   &\qquad +w_{\mathcal{CACACAC}}\mathcal{CACACAC}]_{ii} \label{eq:elementary_decomposition} \\
\nonumber
	d_{*,i} &=[ w_{\mathcal{AAA}} \mathcal{AFA} + w_{\mathcal{AACAC}} \mathcal{AFCAC} + w_{\mathcal{ACAAC}} \mathcal{ACFAC} \\ \nonumber
   &\qquad  +w_{\mathcal{ACACA}} \mathcal{ACFCA} + w_{\mathcal{ACACAC}} \mathcal{ACFCAC}  \\ \nonumber
   &\qquad  +w_{\mathcal{CAAAC}} \mathcal{CAFAC} +  w_{\mathcal{CAACAC}} \mathcal{CAFCAC}  \\
   &\qquad +w_{\mathcal{CACACAC}}\mathcal{CACFCAC}]_{ii} \,, \label{eq:elementary_decomposition_norm}
\end{align}
where $i$ is a node-layer pair and the $w_{\mathcal{E}}$ coefficients are scalars that correspond to the weights for each type of elementary cycle. These weights are different for different types of clustering coefficients; one can choose whatever is appropriate for a given problem. Note that we have absorbed the parameters $\beta$ and $\gamma$ into these coefficients (see below and Table~\ref{table:cc_canonical}). We illustrate the possible elementary cycles in Fig.~\ref{fig:elementary_cycles}
of the main text and in Fig.~\ref{fig:elementary_cycles_others}.

One can even express the cycles that include two consecutive inter-layer steps in the standard form of Eqs.~{\bf\ref{eq:elementary_decomposition}}--{\bf\ref{eq:elementary_decomposition_norm}} for node-aligned multiplex networks, because $\mathcal{C}^2=(b-1)\mathcal{I}+(b-2)\mathcal{C}$ in this case. Without the assumption that $\beta=\gamma=1$, the expansion for the coefficient $c_{SM,i}$ is cumbersome because it includes coefficients $\beta^k\gamma^h$ with all possible combinations of $k$ and $h$ such that $k+h=6$ and $h \neq 1$. Furthermore, in the general case, it is also not possible to infer the number of layers in which a walk traverses an intra-layer edge based on the exponents of $\beta$ and $\gamma$ for $c_{SM,i}$ and $c_{SM^\prime,i}$. For example, in $c_{SM^\prime,i}$, the intra-layer elementary triangle $\mathcal{AAA}$ includes a contribution from both $\beta^3$ (i.e., the walk stays in the original layer) and $\beta\gamma^2$ (i.e., the walk visits some other layer but then comes back to the original layer without traversing any intra-layer edges while it was gone).  Moreover, all of the terms with $b$ arise from a walk moving to a new layer and then coming right back to the original layer in the next step. Because there are $b-1$ other layers from which to choose, the influence of cycles with such transient layer visits is amplified by the total number of layers in a network. That is, adding more layers (even ones that do not contain any edges) changes the relative importance of different types of elementary cycles.

In Table~\ref{table:cc_canonical}, we show the values of the coefficients $w_\mathcal{E}$ for the different ways that we define 3-cycles in multiplex networks.  In Table~\ref{table:cc_canonical_simple}, we show their corresponding expansions in terms of elementary cycles for the case $\beta=\gamma=1$. These cycle decompositions illuminate the difference between $c_{M,i}$, $c_{M^\prime,i}$, $c_{SM,i}$, and $c_{SM^\prime,i}$. The clustering coefficient $c_{M,i}$ gives equal weight to each elementary cycle, whereas $c_{M^\prime,i}$ gives half of the weight to $\mathcal{AAA}$ and $\mathcal{ACACA}$ cycles (i.e., the elementary cycles that include an implicit double-counting) as compared to the other cycles. The matrices that correspond to elementary cycles with such double-counting are symmetric, and the same cycle is thus counted in two different directions.

\begin{sidewaystable}
\vspace{15cm}
\caption{Coefficients of elementary multiplex 3-cycle terms $w_{\mathcal{E}}$ (see Eqs.~{\bf\ref{eq:elementary_decomposition}} and {\bf\ref{eq:elementary_decomposition_norm}}) for different multiplex clustering coefficients. For example, $w_{\mathcal{AAA}}$ for type-$M$ clustering coefficients (i.e., $C_{M}$, $C_{M,u}$, and $c_{M,i}$) is equal to $2\beta^3$. For type-$SM'$ and type-$SM$ clustering coefficients, we calculate the expansions only for node-aligned multiplex networks.
}
\begin{center}
\begin{tabular}{c|c|cccccccc}
  \hline
  C.C.            &$\beta^h\gamma^k$      & $\mathcal{AAA}$ & $\cal {AACAC}$ & $\cal {ACAAC}$ & $\cal {ACACA}$ & $\cal {ACACAC}$ & $\cal {CAAAC}$ & $\cal {CAACAC}$ & $\cal {CACACAC}$ \\ \hline
                  &  $\beta^3$           &   $2$   &        &        &        &         &        &         &          \\
  $M $        &  $\beta\gamma^2$      &      &  $2$      & $2$       & $2$       &         &        &         &          \\
                  &  $\gamma^3$            &      &        &        &        &  $2$       &        &         &          \\
  \hline
                  &  $\beta^3$           &   $1$   &        &        &        &         &        &         &          \\
  $M^\prime $   &  $\beta\gamma^2$      &      &  $2$      & $2$       & $1$       &         &        &         &          \\
                  &  $\gamma^3$            &      &        &        &        &  $2$       &        &         &          \\
  \hline
                  &  $\beta^3$           &  $1$    &        &        &       &         &        &         &          \\
  $SM^{\prime}$  &  $\beta\gamma^2$      &  $2(b-1)$  &  $2$    &   $4$    &    $3$    &         &   $1$    &         &          \\
                  &  $\gamma^3$            &      &        & $2(b-2)$   & $2(b-2)$    &   $2$     &        &    $2$    &          \\
  \hline
  \multirow{6}{*}{$SM$}   &  $\beta^6$           & $1$      &           &           &         &            &         &           &          \\
                  &  $\beta^4\gamma^2$    &$2(b-1)$    &  $4$      &  $4$      &  $4$    &            &  $1$    &           &          \\
                  &  $\beta^3\gamma^3$    &          &$2(b-2)$     &  $2(b-2)$   &  $4(b-2)$ &     $8$    &         & $4$       &          \\
                  &  $\beta^2\gamma^4$    &$(b-1)^2$&  $4(b-1)$   &$4(b-1)$     &$(b-2)^2$&  $8(b-2)$   &  $2(b-1)$ & $2(b-2)$     &  $4$        \\
                  &  $\beta^1\gamma^5$    &          &$2(b-2)(b-1)$&$2(b-2)(b-1)$&         &$2(b-2)^2$ &         &  $4(b-1)$    &   $4(b-2)$       \\
                  &  $\gamma^6$            &          &          &           &         &            &$(b-1)^2$&$2(b-2)(b-1)$ &$(b-2)^2$   \\
  \hline
\end{tabular}
\end{center}
\label{table:cc_canonical}
\end{sidewaystable}

\begin{sidewaystable}[htp]
\vspace{15cm}
\caption{Coefficients of the elementary multiplex 3-cycle terms $w_{\mathcal{E}}$ (see Eqs.~{\bf\ref{eq:elementary_decomposition}} and {\bf\ref{eq:elementary_decomposition_norm}}) for different multiplex clustering coefficients when $\beta=\gamma=1$. For type-$SM'$ and type-$SM$ clustering coefficients, we calculate the expansions only for node-aligned multiplex networks.
}
\begin{center}
\begin{tabular}{l|c|ccc|c|ccc}
  \hline
  C. C.           & $\cal {AAA}$ & $\cal {AACAC}$ & $\cal {ACAAC}$ & $\cal {ACACA}$ & $\cal {ACACAC}$ & $\cal {CAAAC}$ & $\cal {CAACAC}$ & $\cal {CACACAC}$ \\ \hline
  $M$         & $2$   & $2$     & $2$     & $2$     & $2$      & $0$     & $0$      & $0$       \\
  $M^\prime$    & $1$   & $2$     & $2$     & $1$     & $2$      & $0$     & $0$      & $0$       \\
  $SM$        & $1b^2$ & $2b^2$  & $2b^2$  & $b^2$   & $2b^2$    & $b^2$   & $2b^2$   & $b^2$      \\
  $SM'$       & $2b-1$ & $2$     & $2b$   & $2b-1$   & $2$      & $1$     & $2$      & $0$      \\
  \hline
\end{tabular}
\end{center}
\label{table:cc_canonical_simple}
\end{sidewaystable}





\newpage

\section{A Simple Example}
\label{sec:example}

\begin{figure}
  \begin{center}
    \includegraphics[width=0.25\textwidth]{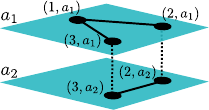}
  \end{center}
  \caption{A simple, illustrative example of a multiplex network.}
  \label{fig:simple_ex}
\end{figure}

We now use a simple example (see Fig.~\ref{fig:simple_ex}) to illustrate the differences between the various notions of a multiplex clustering coefficient. Consider a two-layer multiplex network with three nodes in layer $a_1$ and two nodes in layer $a_2$. The three node-layer pairs in layer $a_1$ form a 2-star, the two node-layer pairs that are not connected directly to each other on layer $a_1$ are each adjacent via an inter-layer edge to a counterpart node-layer pair in layer $a_2$, and the two node-layer pairs on layer $a_2$ are adjacent to each other.

The adjacency matrix $\mathcal{A}$ for the intra-layer supra-graph is
\begin{equation}
	\mathcal{A}=
 \left( \begin{array}{ccccc}
0 & 1 & 1 & 0 & 0 \\
1 & 0 & 0 & 0 & 0\\
1 & 0 & 0 & 0 & 0\\
0 & 0 & 0 & 0 & 1\\
0 & 0 & 0 & 1 & 0\end{array} \right)\,,
\end{equation}
and the adjacency matrix $\mathcal{C}$ of the coupling supra-graph is
\begin{equation}
	\mathcal{C}=
 \left( \begin{array}{ccccc}
0 & 0 & 0 & 0 & 0 \\
0 & 0 & 0 & 1 & 0\\
0 & 0 & 0 & 0 & 1\\
0 & 1 & 0 & 0 & 0\\
0 & 0 & 1 & 0 & 0\end{array} \right)\,.
\end{equation}
Therefore, the supra-adjacency matrix is
\begin{equation}
	\mathcal{\bar{A}}=
 \left( \begin{array}{ccccc}
0 & 1 & 1 & 0 & 0 \\
1 & 0 & 0 & 1 & 0\\
1 & 0 & 0 & 0 & 1\\
0 & 1 & 0 & 0 & 1\\
0 & 0 & 1 & 1 & 0\end{array} \right)\,.
\end{equation}
The multiplex walk matrix ${\cal M}_{M}$ is
\begin{equation}
	\mathcal{M}_{M}=\sqrt[3]{2}
 \left( \begin{array}{ccccc}
0 & 1 & 1 & 1 & 1 \\
1 & 0 & 0 & 0 & 0\\
1 & 0 & 0 & 0 & 0\\
0 & 0 & 1 & 0 & 1\\
0 & 1 & 0 & 1 & 0\end{array} \right),
\end{equation}
and we note that it is not symmetric.  For example, node-layer pair $(2,a_2)$ is reachable from $(1,a_1)$, but node-layer pair $(1,a_1)$ is not reachable from $(2,a_2)$. The edge $[(1,a_1),(2,a_2)]$ in this supra-graph represents the walk $\{(1,a_1),(2,a_1),(2,a_2)\}$ in the multiplex network. The symmetric walk matrix $\mathcal{M}_{SM^\prime}$ is
\begin{equation}
	\mathcal{M}_{SM^\prime}=
 \left( \begin{array}{ccccc}
0 & 1 & 1 & 1 & 1 \\
1 & 0 & 0 & 0 & 1\\
1 & 0 & 0 & 1 & 0\\
1 & 0 & 1 & 0 & 1\\
1 & 1 & 0 & 1 & 0\end{array} \right)\,.
\end{equation}
The matrix $\mathcal{M}_{SM^\prime}$ is the sum of $\mathcal{M}_{M}$  and $\mathcal{M}_{M}^T$ with rescaled diagonal blocks in order to not double-count the edges $[(1,a_1),(2,a_1)]$ and $[(1,a_1),(3,a_1)]$. Additionally,
\begin{equation}
	\mathcal{M}_{SM}=
 \left( \begin{array}{ccccc}
0 & 1 & 1 & 1 & 1 \\
1 & 0 & 1 & 0 & 1\\
1 & 1 & 0 & 1 & 0\\
1 & 0 & 1 & 0 & 1\\
1 & 1 & 0 & 1 & 0\end{array} \right)\,,
\end{equation}
which differs from $\mathcal{M}_{SM^\prime}$ in the fact that node-layer pairs $(2,a_1)$ and $(3,a_1)$ are connected through the multiplex walk $\{(2,a_1),(2,a_2),(3,a_2),(3,a_1)\}$.

The adjacency matrix of the aggregated graph is
\begin{equation}
	{\bf W}=
 \left( \begin{array}{ccc}
0 & 1 & 1\\
1 & 0 & 1\\
1 & 1 & 0
\end{array} \right)\,.
\end{equation}
That is, it is a complete graph without self-edges.

We now calculate $c_{*,i}$ using different definitions of a multiplex clustering coefficient. To calculate $c_{M,i}$, we need to compute the auxiliary complete supra-adjacency matrix $\mathcal{M}_{M}^F$ according to Eq.~{\bf\ref{eq20}}. We obtain
\begin{equation}
	\mathcal{M}_{M}^F=\sqrt[3]{2}\mathcal{F\widehat{C}}=\sqrt[3]{2}
 \left( \begin{array}{ccccc}
0 & 1 & 1 & 1 & 1 \\
1 & 0 & 1 & 0 & 1\\
1 & 1 & 0 & 1 & 0\\
0 & 0 & 1 & 0 & 1\\
0 & 1 & 0 & 1 & 0\end{array} \right)\,.
\end{equation}
The clustering coefficient of node-layer pair $(1,a_1)$, which is part of two triangles that are reachable
 along the directions of the edges, is
\begin{equation}
	c_{M,(1,a_1)}=\frac{1}{2}\,.
\end{equation}
For node-layer pair $(2,a_1)$, we get
\begin{equation}
	c_{M,(2,a_1)}=1\,,
\end{equation}
which is the same as the clustering-coefficient values of the remaining node-layer pairs.

To calculate $c_{SM^\prime,i}$, we need to compute $\cal {M}_{SM^\prime}^F$, which we obtain using Eq.~{\bf\ref{sawprime}}. We thus obtain
 \begin{equation}
	\mathcal{M}_{SM^\prime}^F=\mathcal{F\widehat{C}}+\mathcal{\widehat{C}F}=
 \left( \begin{array}{ccccc}
0 & 1 & 1 & 1 & 1 \\
1 & 0 & 1 & 0 & 1\\
1 & 1 & 0 & 1 & 0\\
1 & 0 & 1 & 0 & 1\\
1 & 1 & 0 & 1 & 0\end{array} \right)\,.
\end{equation}
In the supra-graph associated with the supra-adjacency matrix $\mathcal{F\widehat{C}}+\mathcal{\widehat{C}F}$, all node-layer pairs are adjacent to all other node-layer pairs except those that correspond to the same physical nodes. 
The clustering coefficient of node-layer pair $(1,a_1)$, which is part of six triangles, is
\begin{equation}
	c_{SM^\prime,(1,a_1)}=\frac{1}{2}=c_{M,(1,a_1)}\,.
\end{equation}
The clustering coefficient of node-layer pair $(2,a_1)$, which is part of one triangle, is
\begin{equation}
	c_{SM^\prime,(2,a_1)}=1\,.
\end{equation}

To calculate $c_{SM,i}$, we compute $\mathcal{M}_{SM}^F$ using Eq.~{\bf\ref{saw}}. We thus obtain
 \begin{equation}
	\mathcal{M}_{SM}^F=\mathcal{\widehat{C}F\widehat{C}}=
 \left( \begin{array}{ccccc}
0 & 1 & 1 & 1 & 1 \\
1 & 0 & 2 & 0 & 2\\
1 & 2 & 0 & 2 & 0\\
1 & 0 & 2 & 0 & 2\\
1 & 2 & 0 & 2 & 0\end{array} \right)\,.
\end{equation}
The only difference between the graphs associated with the matrices $\mathcal{\widehat{C}F\widehat{C}}$ and $\mathcal{F\widehat{C}}+\mathcal{\widehat{C}F}$ is the weight of the edges in $\mathcal{\widehat{C}F\widehat{C}}$ that take into account the fact that intra-layer edges might be repeated in the two layers.

The clustering coefficient of node-layer pair $(1,a_1)$, which is part of eight triangles, is
\begin{equation}
	c_{SM,(1,a_1)}=\frac{8}{12}=\frac{2}{3}\,.
\end{equation}
The clustering coefficient of node-layer pair $(2,a_1)$, which is part of four triangles, is
\begin{equation}
	c_{SM,(2,a_1)}=\frac{4}{6}=\frac{2}{3}\,.
\end{equation}

Because we are weighting edges based on the number of times an edge between two nodes is repeated in different layers among a given pair of physical nodes in the normalization, none of the node-layer pairs has a clustering coefficient equal to $1$. By contrast, all nodes have clustering coefficients with the same value in the aggregated network, for which the layer information has been lost. In particular, they each have a clustering-coefficient value of $1$, independent of the definition of the multiplex clustering coefficient.


\section{Further Discussion of Clustering Coefficients in Multiplex Erd\H{o}s-R\'{e}nyi (ER)  Networks}
\label{sec:ercc2}

The expected values of the local clustering coefficients in node-aligned multiplex ER networks are
\begin{align}
	\ensavg{c_{\mathcal{AAA},\micronode}}&=\frac{1}{b}\sum_{\alpha \in L} p_{\alpha} \equiv \avg{p}\,, \\
\ensavg{c_{\mathcal{AACAC},\micronode}}&=\frac{1}{b}\sum_{\alpha \in L} p_{\alpha} \equiv \avg{p}\,, \\
\ensavg{c_{\mathcal{ACAAC},\micronode}}&=\frac{1}{b}\sum_{\alpha \in L}\frac{\sum_{\kappa\neq\alpha}p_{\kappa}^2}{\sum_{\kappa\neq\alpha}p_{\kappa}}\,, \\
	\ensavg{c_{\mathcal{ACACA},\micronode}}&=\frac{1}{b}\sum_{\alpha \in L} p_{\alpha} \equiv \avg{p}\,, \\
\ensavg{c_{\mathcal{ACACAC},\micronode}}&=\frac{1}{b(b-1)}\sum_{\alpha \in L}\frac{\sum_{\kappa\neq\alpha;\mu\neq\kappa,\alpha}p_{\kappa} p_{\mu}}{\sum_{\kappa\neq\alpha}p_{\kappa}}.
\end{align}
Note that $c_{M,\micronode}^{(1)}=c_{\mathcal{AAA},\micronode}$ and $c_{M,\micronode}^{(3)}=c_{\mathcal{ACACAC},\micronode}$, but the 2-layer clustering coefficient $c_{M,\micronode}^{(2)}$ arises from a weighted sum of contributions from three different elementary cycles.

\begin{figure}[t]
  \begin{center}
    \includegraphics[width=.45\textwidth]{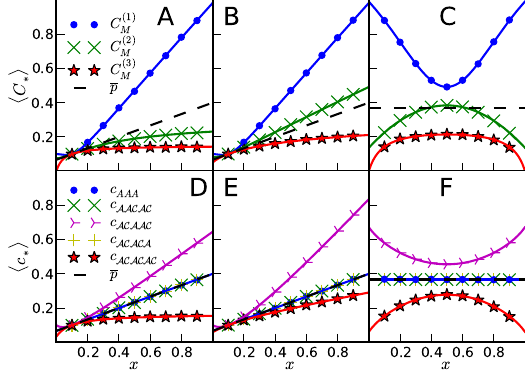}
  \end{center}
  \caption{(A, B, C) Global and (D, E, F) local multiplex clustering coefficients in multiplex networks that consist of ER layers. The markers give the results of simulations of 100-node node-aligned multiplex ER networks that we average over 10 realizations. The solid curves are theoretical approximations (see Eqs.~{\bf\ref{12}}--{\bf \ref{14}} of the main text). Panels (A, C, D, F) show results for three-layer networks, and panels (B, E) show results for six-layer networks. The ER edge probabilities of the layers are (A, D) $\{0.1, 0.1, x\}$, (B, E) $\{0.1, 0.1, 0.1, 0.1, x, x\}$, and (C, F) $\{0.1, x, 1-x\}$.}
\label{fig:er_global_local}
\end{figure}

In Fig.~\ref{fig:er_global_local}, we illustrate the behavior of the global and local clustering coefficients in multiplex networks in which the layers consist of ER networks with varying amounts of heterogeneity in the intra-layer edge densities. Although the global and mean local clustering coefficients are equal to each other when averaged over ensembles of monoplex ER networks, we do not obtain a similar result for multiplex networks with ER layers unless the layers have the same value of the parameter $p$. The global clustering coefficients give more weight than the mean local clustering coefficients to denser layers. This is evident for the intra-layer clustering coefficients $c_{M,\micronode}^{(1)}$ and $C_{M}^{(1)}$, for which the ensemble average of the mean of the local clustering coefficient $c_{M,\micronode}^{(1)}$ is always equal to the mean edge density, whereas the ensemble average of the global clustering coefficient $C_{M}^{(1)}$ has values that are greater than or equal to the mean edge density. This situation is a good example of a case in which transitivity in multiplex networks differs from the results and intuition from monoplex networks.

In particular, failing to take into account the heterogeneity of edge densities in multiplex networks can lead to incorrect or misleading results when trying to distinguish among values of a clustering coefficient that are what one would expect from an ER random network versus those that are a signature of a triadic-closure process (see Fig.~\ref{fig:er_global_local}).


\section{Null Model for Shuffling Inter-layer Connections}
\label{sec:interlayershuffle}

In Table~\ref{table:cc_valuesH}, we compare empirical values of layer-decomposed global clustering coefficients with clustering-coefficient values for a null model in which we preserve the topology of each intra-layer network but independently
shuffle the labels of the nodes inside of each layer.
 That is, for each intra-layer network $G^\alpha=(V^\alpha,E^\alpha)$, we 
choose a permutation $\pi: V^\alpha \mapsto V^\alpha$ uniformly at random and construct a new multiplex network starting from $\{ \pi(G^\alpha) \}$, where $\pi(G^\alpha)=(\pi(V^\alpha),\pi(E^\alpha))$ and $\pi(E^\alpha)=\{ (\pi(u),\pi(v)) | (u,v) \in E^\alpha \}$. In this way, we effectively randomize inter-layer edges but preserve both the structure of intra-layer networks and the number of inter-layer edges between each pair of layers. For our comparisons using this null model, most of the clustering coefficients take values that are significant for our data sets 
(see Table~\ref{table:cc_valuesH}). Because of the way that we construct the null model, the global single-layer clustering coefficients are exactly the same for the original data and the null model.

\begin{sidewaystable}
\vspace{15cm}
\begin{tabular}{lccccccc}
\multicolumn{2}{l}{}  &Tailor Shop&Management&Families&Bank&Tube&Airline\\
\hline
\multirow{2}{*}{$C_{M} $} &orig.&0.319**&0.206**&0.223&0.293**&0.056**&0.101** \\
&NM&$0.218\pm 0.007$&$0.131\pm 0.003$&$0.194\pm 0.029$&$0.223\pm 0.014$&$0.025\pm 0.008$&$0.054\pm 0.001$ \\
\hline
\multirow{2}{*}{$C_{M}^{(1)}$} &orig.&0.406&0.436&0.289&0.537&0.013&0.100 \\
&NM&$0.406$&$0.436$&$0.289$&$0.537$&$0.013$&0.100 \\
\hline
\multirow{2}{*}{$C_{M}^{(2)}$} &orig.&0.327**&0.273**&0.198&0.349**&0.043&0.150** \\
&NM&$0.220\pm 0.009$&$0.176\pm 0.003$&$0.158\pm 0.041$&$0.240\pm 0.018$&$0.035\pm 0.018$&$0.082\pm 0.002$ \\
\hline
\multirow{2}{*}{$C_{M}^{(3)}$} &orig.&0.288**&0.192**&-&0.227'&0.314**&0.086** \\
&NM&$0.165\pm 0.010$&$0.120\pm 0.003$&-&$0.186\pm 0.016$&$0.053\pm 0.043$&$0.037\pm 0.002$ \\
\end{tabular}
\caption{
Clustering coefficients $C_{M}$, $C_{M}^{(1)}$, $C_{M}^{(2)}$, $C_{M}^{(3)}$ that correspond, respectively, to the global, one-layer, two-layer, and three-layer clustering coefficients for various multiplex networks. ``Tailor Shop'': Kapferer tailor-shop network ($n=39$, $b=4$)~\cite{Kapferer1972Strategy}. ``Management'': Krackhardt office cognitive social structure ($n=21$, $b=21$)~\cite{Krackhardt1987Cognitive}. ``Families'': Padgett Florentine families social network ($n=16$, $b=2$)~\cite{Breiger1986Cumulated}. ``Bank'': Roethlisberger and Dickson bank wiring-room social network ($n=14$, $b=6$)~\cite{Roethlisberger1939Management}. ``Tube'': The London Underground (i.e., ``The Tube'') transportation network ($n=314$, $b=14$)~\cite{Rombach2012Core}. ``Airline'': Network of flights between cities, in which each layer corresponds to a single airline ($n=3108$, $b=530$)~\cite{airlinedata}.
The rows labeled ``orig.'' give the clustering coefficients for the original networks, and the rows labeled
``NM'' give the expected value and the standard deviation of the clustering coefficient in a null-model network in which we preserve the topology of the intra-layer networks but separately (and independently) shuffle the node labels for each intra-layer network. For the original values, we perform a two-tailed Z-test to examine whether the observed clustering coefficients could have been produced by the null model. We designate the p-values as follows: *: $p<0.05$, **: $p<0.01$ for Bonferroni-corrected tests with 18 hypothesis; ': $p<0.05$, '': $p<0.01$ for uncorrected tests. We do not use any symbols for values that are not significant. We symmetrize directed networks by considering two nodes to be adjacent if there is at least one edge between them. The social networks in this table are node-aligned multiplex graphs, but the transportation networks are not node-aligned. We report values that are means over different numbers of realizations: $10^5$ for Tailor Shop, $10^3$ for Airline, $10^4$ for Management, $10^5$ for Families, $10^4$ for Tube, and $10^5$ for Bank. 
}
\label{table:cc_valuesH}
\end{sidewaystable}






\clearpage
\vspace{1cm}


\providecommand{\newblock}{}

\end{document}